\documentclass{article}
\usepackage{arxiv}

\usepackage{lineno,hyperref}
\modulolinenumbers[5]

\usepackage{amssymb}

\usepackage[figuresright]{rotating}
\usepackage[utf8]{inputenc} 
\usepackage[T1]{fontenc}    
\usepackage{hyperref}       
\usepackage{url}            
\usepackage{booktabs}       
\usepackage{amsfonts}       
\usepackage{nicefrac}       
\usepackage{microtype}      
\usepackage{xcolor}
\usepackage{graphicx}
\usepackage{amssymb}
\usepackage{amsmath}
\usepackage{enumitem}
\usepackage{algorithmic}
\usepackage[ruled]{algorithm2e}
\usepackage{siunitx}
\usepackage{caption}
\usepackage{subcaption}
\usepackage[super]{nth}
\usepackage[normalem]{ulem}

\newcommand{\eqr}[1]{$#1$}

\title{Self-play Learning Strategies for Resource Assignment in Open-RAN Networks}

\author{Xiaoyang Wang\;\;\;\;\;\;\;\;\;\;\;\;\;\;\;\;Jonathan D.~Thomas\;\;\;\;\;\;\;\;\;\;\;\;Robert J. Piechocki\;\;\;\;\\
University of Bristol\;\;\;\;\;\;\;\;\;\;\;\;\;\;University of Bristol\;\;\;\;\;\;\;\;\;\;\;\;\;\;University of Bristol\;\;\;\;\;\;\\
\;\;\;\;\;\;\;\;\;\;\;\;\;\;\;\;\;\;\;\;\;\;\;\;\;\;\;\;\;\;\;\;\;\;\;\;\;\;\;\;\;\;\;\;\;\;\;\;\;\;\;\;\;\;\;\;\;\;\;\;\;\;\;\;\;\;\;\;\;\;\;\;\;\;\;\;\;\;\;\;\;\;\;The Alan Turing Institute\\
\texttt{\texttt{\{xiaoyang.wang, jonathan.david.thomas, r.j.piechocki\}@bristol.ac.uk}}\\
\AND
Shipra Kapoor\\
BT\\
\texttt{shipra.kapoor@bt.com}\\
\And
Ra\'ul Santos-Rodr\'iguez\\
University of Bristol\\
\texttt{enrsr@bristol.ac.uk}\\
\And
Arjun Parekh\\
BT\\
\texttt{arjun.parekh@bt.com}
}

\begin{document}
\maketitle

\begin{abstract}
Open Radio Access Network (ORAN) is being developed with an aim to democratise access and lower the cost of future mobile data networks, supporting network services with various QoS requirements, such as massive IoT and URLLC. In ORAN, network functionality is dis-aggregated into remote units (RUs), distributed units (DUs) and central units (CUs), which allows flexible software on Commercial-Off-The-Shelf (COTS) deployments. Furthermore, the mapping of variable RU requirements to local mobile edge computing centres for future centralized processing would significantly reduce the power consumption in cellular networks. In this paper, we study the RU-DU resource assignment problem in an ORAN system, modelled as a 2D bin packing problem. A deep reinforcement learning-based self-play approach is proposed to achieve efficient RU-DU resource management, with AlphaGo Zero inspired neural Monte-Carlo Tree Search (MCTS). Experiments on representative 2D bin packing environment and real sites data show that the self-play learning strategy achieves intelligent RU-DU resource assignment for different network conditions. 
\end{abstract}

\keywords{Open-RAN \and Deep reinforcement learning \and Self-play \and Resource assignment}


\section{Introduction}

Next generation networks promise always-on connectivity everywhere, ultra-low latency and massive capacity. It is seen as a key enabler of the future service revolution \cite{Nokiawhitepaper}. Wireless services have evolved in various directions during the past few decades, with the emerging ecosystem of the Internet of Things (IoT), Ultra Reliable Low Latency Communication (URLLC), high-speed services, etc~\cite{singh2020evolution, ren2020resource, hussain2020machine}. To fully utilise the potential of 5G to provide for services with different Quality of Service (QoS) requirements, a versatile structure with reconfigurability is essential. The existing approach of utilising proprietary equipment and vendor-specific design locks mobile network operators (MNOs) into a specific architecture, making upgrade and interoperability of advanced features difficult~\cite{samsungwhitepaper}. Software-oriented and application-adaptive network architecture will be able to serve massive IoT devices and other co-existing network services. Within a 5G Network, the RAN represents significant expenditure, which is estimated to account for 60-65\% of the total expense of network ownership~\cite{oranwhitepaper}. To deliver 5G and beyond services with cost-effective continuous technology upgrades, Open-RAN (ORAN) emerges as the most promising solution.





The concept of ORAN is a vendor-neutral disaggregated network structure which disengages the software from hardware and vendor. This concept was introduced by the 3rd Generation Partnership Project (3GPP) in Release 14 specifications~\cite{3gppr14}. 3GPP introduced the decomposition of the existing baseband unit (BBU) into three elements, where these are the remote unit (RU), distributed unit (DU) and central unit (CU). In 3GPP release 15 and subsequent specifications, disaggregation continued in service-based architecture with the split between control and user plane in 5G base stations. A disaggregated network structure has several advantages including: {\em (i)} The RAN disaggregation brings higher network utilization efficiency~\cite{yi2018comprehensiveNFV, wu2015cloud}; {\em(ii)} it will enhance network optimisation and improve network quality of experience (QoE) in dynamic environments~\cite{parallelwirelesswhitepaper}; and {\em (iii)} for the RU-DU split, centralized processing across multiple RUs would reduce the cost of baseband resources. 

In an ORAN system, the DU is typically hosted in an edge cloud data centre. The connection between DU and RU is one-to-many. Intelligent DU resource allocation for various RU requirements would significantly improve the utilisation of DU resource, facilitating the promised benefits of centralised processing, such as reduced cost and improved user satisfaction. Resource assignment in the RAN has previously been studied along with the development of network centralisation. \cite{tang2019systematic} studies the resource segmentation and allocation problem in Cloud-RAN (C-RAN), considering physical resource allocation and the time resource. The problem is formulated as a stochastic mixed integer nonlinear programming and is solved by successive convex approximation. Heuristics methods are also studied in RAN resource allocation~\cite{yang2020survivableheu, xia2019programmable}. \cite{mharsi2019cloud} studies both exact and heuristics methods for the edge computing optimization in C-RAN, including integer linear programming, Matroid-based method and knapsack-based method where they jointly optimize the resource consumption and network latency.

In this paper, we model the RU-DU resource assignment problem in ORAN as a bin packing problem, proposing a self-play reinforcement learning strategy. We apply the combined approach of deep neural network with Monte-Carlo Tree Search (MCTS)~\cite{silver2017masteringAlphazero} for solving this combinatorial optimization problem. With high capacity and function approximation ability, a deep neural network modelled resource assignment policy can generalize across dynamic network conditions. Learning from self-play eliminates the need for demonstrator data, which could be expensive and time-consuming to collect.   

The rest of this paper is organised as follows. In Section~\ref{sec: related work}, we review the application of Reinforcement Learning (RL) in next generation networks. Section~\ref{sec: method} presents the RU-DU resource assignment problem studied in this paper and the proposed approach. Experiments on the representative bin packing environment and real scenarios are conducted in Section~\ref{sec: experiments}. In Section~\ref{sec: discussion}, we present the conclusion and future challenges for RL in the ORAN system.

\section{Related Work}
\label{sec: related work}
\subsection{RL in Next Generation Networks}

Due to its proven capability in intelligent decision making~\cite{silver2017masteringchess, mnih2015humanlevelcontrol, vinyals2019grandmasterSCII}, RL has been applied to address communications and networking issues including traffic routing, data offloading, resource allocation, etc~\cite{luong2019applications}. In this section, we focus on RL for resource allocation in 5G and beyond. \cite{zhao2018deepD3QN} proposes a dueling-double DQN approach for large scale user association and resource allocation in heterogeneous cellular networks. In mobile edge computing, RL is widely applied for intelligent edge resource allocation~\cite{liu2019deepMEC, wang2019smartMEC}. A variety of services with diverse requirements are supposed to be supported by network slicing in next generation networks, which can also be achieved by RL approaches~\cite{qi2019deepslicing, wang2019dataslicing}. Spectrum allocation and sharing approaches are studied in~\cite{liang2019spectrum, lin2020dynamicspectrum}, improving the link capacity by RL methods. 

\subsection{RL for Combinatorial Optimisation}
The combinatorial optimization (CO) problem can be viewed as searching for the optimal solution in the feasible region~\cite{bengio2020machinelearning}. Typical CO problems include the Travelling Salesman Problem (TSP), Bin Packing Problem (BPP), Vehicle Routing Problem (VRP), etc. In recent years, RL approaches have been designed to solve CO problems, by learning policies to either incrementally build CO solutions or iteratively improve solutions~\cite{mazyavkina2020reinforcementsurvey}. In this section, we especially focus on the neural MCTS method, i.e., the AlphaGo Zero method, reviewing its application in combinatorial optimization problems. AlphaGo Zero is a model-based method, in which the model is given for the agent to plan ahead through MCTS. It belongs to the paradigm of joint approaches, where the decision for CO problems is made by RL policy guided MCTS search. \cite{laterre2018ranked} proposes to use a reward-shaping mechanism called ranked reward, enabling single-player games to benefit from adversarial games settings. This approach successfully solved 2D and 3D BPP with a reasonable scale. \cite{wang2020tacklingMS} applied the ranked reward neural MCTS approach to the Morpion Solitaire game, resulting in the best non-human performance. In~\cite{xu2019learning}, the CO problems are transformed into the Zermelo game domain, from which optimal solutions can be learned using neural MCTS. CO problem on graphs is studied in~\cite{abe2019solvingNPHard}, in which a graph isomorphism network with MCTS is proposed without human knowledge.

\section{Methodology}
\label{sec: method}
In this section, we introduce the RU-DU resource assignment in an ORAN system as a variation of the classic bin packing problem. We formulate a 2D bin packing problem to pack items into a single bin while minimizing one dimension of the bin. A self-play reinforcement learning approach is proposed to solve the 2D bin packing problem.
\subsection{RU-DU Resource Assignment in ORAN}
\label{sec:RU-DU 2D bin}
\begin{figure}[htp]
    \centering
    \includegraphics[width=3 in]{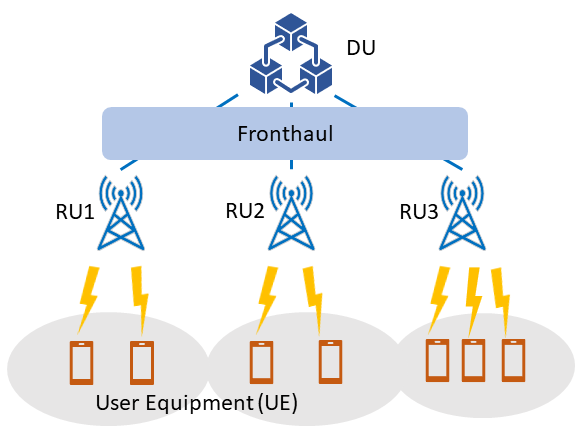}
    \caption{RU-DU architecture in ORAN.}
    \label{fig:RU-DU}
\end{figure}

We consider the edge-central processing model in Fig.~\ref{fig:RU-DU}. The DU is colocated with multiple RUs, connected through fronthaul interfaces. According to the 5G New Radio (NR) functional split design, RUs are small and cost-effective, while DUs are designed to undertake the role of computing centres, conducting the majority of data processing. Resource-intense jobs are centralised at DUs~\cite{larsen2018surveysplit}. Intelligent resource assignment between DU and RU will result in high utilisation of DU resources while satisfying latency requirements, bringing the benefit of cost-efficient network operations.

During network operation, the~\eqr{i^\text{th}} RU, represented as \eqr{R_i}, requires compute resource for radio signal processing. \eqr{R_i} raises requests to the DU for resources with capacity \eqr{c_i} and an estimated processing time \eqr{t_i}. These requests are driven by its connection status with UEs, as well as the UE demands. DUs, as mobile edge computing centres, are modelled as resource clusters with capacity \eqr{C}. RU requests are allocated and processed on a batch basis. Note that in this work, we assume the DU-RU connection status is known. We only focus on the resource assignment in established one-to-many DU-RU connections. This problem can be modeled as a 2D bin packing problem (BPP), where RU requirements are modelled as a batch of rectangular items \eqr{\mathcal{I} = \{w_i, h_i\}_{i=1}^N} and DU is modelled as a bin of size \eqr{(\tilde{W}, \tilde{H})}. Here \eqr{w_i = t_i}, \eqr{h_i =c_i}, and \eqr{N} is the number of RUs connected to one DU. For the bin (the DU),  \eqr{\tilde{W}} is the time limit for processing the current batch of RU requirements, and \eqr{\tilde{H}} is the resource capacity. RU-DU resource assignment is to find the placement of items in the bin. We define~\eqr{(x_i, y_i)} as the left-bottom coordinate of a placed item~\eqr{i}, with~\eqr{(0, 0)} as the left-bottom coordinate of the bin.

Since the cost of DU cluster construction and maintenance is proportional to the amount of resource it has/uses over the operating time, network operators are looking for intelligent assignment strategies where RUs use the minimum resources at DUs with latency requirements satisfied. Minimising the resource assigned for RUs in a DU will result in free physical resources. They can either be set to sleep mode to save energy, or to support other types of services and open interface operations. Given the latency requirement for a batch of RU requirements \eqr{W^*}, the objective is to find a bin with minimal height where all items can be packed into. This can be formalized as the following optimisation problem~\footnote{The objective is to minimize the height of the bin since it represents the physical resource. Switching the meaning of height and width of the bin does not affect the results in this paper.}:

\begin{align}
    \min \; & \tilde{H} \\
     \textrm{s.t.} \;\; & \tilde{W} = W^* \\
            & \beta_{i,j} + \gamma_{i,j} = 1\\
            &  0 \leq x_i \leq \tilde{W} - w_i \;\;\;\;\;\;\;\;\;\;\;\;\;\;\;\;\;,\forall i \in [1, N]~\label{eq: inside_x}\\
            &  0 \leq y_i \leq \tilde{H} - h_i~\label{eq:inside_y}\\
            &  x_i - x_j + \tilde{W} \cdot \beta_{i,j} \leq \tilde{W}-w_i~\label{eq: overlap_x}\\
            &  y_i - y_j + \tilde{H} \cdot \gamma_{i,j} \leq \tilde{H}-h_i~\label{eq: overlap_y}
\end{align}
Here, \eqr{\beta_{i,j}} and~\eqr{\gamma_{i,j}} are binary values. \eqr{\beta_{i,j}=1} if~\eqr{x_i \leq x_j}, \eqr{\gamma_{i,j}=1} if ~\eqr{y_i \leq y_j}. \eqref{eq: inside_x} and \eqref{eq:inside_y} are to assure items are placed inside the bin. \eqref{eq: overlap_x} and \eqref{eq: overlap_y} guarantee there is no overlap between items, due to the nature of resource and time occupation. In addition, we introduce three placement rules on the 2D BPP:
\begin{itemize}[noitemsep, topsep=0pt]
    \item Items are non-rotatable, as axes are not interchangeable.
    \item An item can only be placed at the edge of the bin or adjacent to another item in both dimensions. This reduces the size of searching space in this problem.
    \item If not specified otherwise, the resource occupancy does not need to be contiguous for each item, as long as the required amount of resources are reserved for the same time period. 
    \item Once placed, an item is not allowed to be removed from the bin or to conduct any kind of re-placement. This is to ensure the item placement process is time-finite, helping to find feasible solutions. 
\end{itemize}

Note that in this work we only consider a single-dimensional resource in mobile edge computing, which already brings benefits to ORAN systems. The challenge of multi-dimensional resource assignment is discussed in Section~\ref{sec: discussion}.

\subsection{Markov Decision Process}
\begin{figure}[ht]
    \centering
    \includegraphics[width=3.2in]{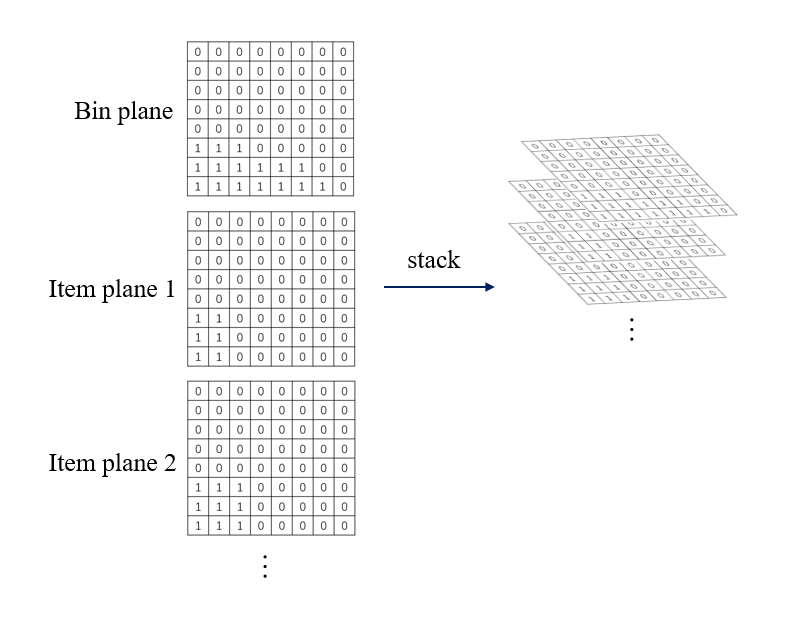}
    \caption{State representation of 2D BPP.}
    \label{fig:state_space}
\end{figure}

\begin{figure*}[htb]
    \centering
    \begin{subfigure}[c]{0.35\textwidth}
        \includegraphics[width = 1.8 in]{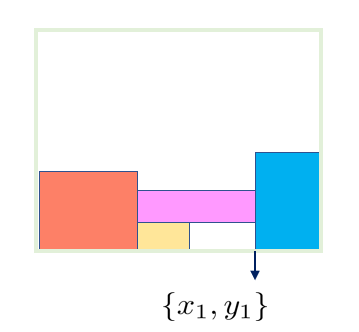}
        \caption{Contiguous resource assignment.}
        \label{fig:contiguous_resource}
    \end{subfigure}
    \hspace{0.6 cm}
    \begin{subfigure}[c]{0.35\textwidth}
        \includegraphics[width = 1.8 in]{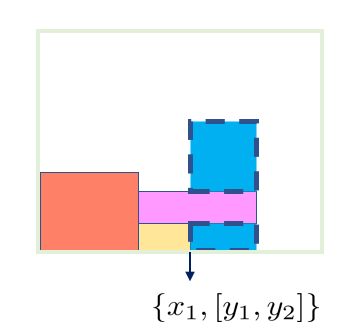}
        \caption{Non-contiguous resource assignment.}
        \label{fig:non-contiguous resource}
    \end{subfigure}
\caption{Item placement for contiguous and non-contiguous resource assignment. For contiguous resource assignment, the placement of the \textit{blue} item is~\eqr{\{x_1, y_1\}}, which is its bottom-left coordination. For non-contiguous resource assignment, the \textit{blue} item can be sliced horizontally to two smaller items, placed in \eqr{\{x_1, y_1\}} and \eqr{\{x_1, y_2\}}, respectively. Thus \eqr{y} is a vector rather than a single value. In this paper, we apply the non-contiguous resource assignment principle, although not all items will be sliced to smaller ones. For consistency, the y-axis placement of item \eqr{i} with size \eqr{w_i, h_i} is written as \eqr{y_i = [y^1, y^2,...,y^{h_i}]}.}
\label{fig:action_fig}
\end{figure*}

We formulate the 2D BPP as a finite Markov decision process (MDP). The bin and items are represented as 2D binary occupancy planes with identical size. For the bin plane, it shows the current placement status. For item planes, unpacked items are placed at the bottom left of the plane while packed items are removed from corresponding planes. The bin plane and item planes are stacked together to form the state of an MDP (see Fig.~\ref{fig:state_space}).
With one item to be placed into the bin at each step, \eqr{\{i, x_i, y_i\}_t,\;t=[1,...,N]} encodes the placement location of item \eqr{i} at step \eqr{t}. Considering the possibly non-contiguous resource assignment for each item, \eqr{y_i = [y^1, y^2,...,y^{h_i}]} is a set of \eqr{h_i} locations rather than a single location (see Fig.~\ref{fig:action_fig}). For simplicity, we define action \eqr{a_t \in \mathcal{A}} as a tuple \eqr{a_t := <i, x_i>_t}. \eqr{y_i} is then determined by a simple heuristic method introduced in Section~\ref{sec: r2 algorithm}. This process is Markovian. The reward is defined as:
\begin{equation}
    r = 
\begin{cases}
    \frac{H^*}{\tilde{H}} & \text{if} \;\; s_t = s^{*} \\
  0 & \text{otherwise}
\end{cases}
\end{equation}
Here \eqr{s^*} is the terminal state of the 2D BPP with all items placed in the bin. \eqr{H^* = \max ( \frac{\sum_{i=1}^N w_i * h_i}{W^*}, \max_i{h_i})} is the possible minimal height of the bin. \eqr{\tilde{W}} and \eqr{\tilde{H}} are the width and height of the bin at the terminal state. Thus, \eqr{r \in (0, 1]}.

\subsection{Ranked Reward with Self-play Reinforcement Learning}
\label{sec: r2 algorithm}
As described in Section~\ref{sec:RU-DU 2D bin}, RU requests are processed in batches. Each batch of RU requests forms an instance of the 2D BPP with \eqr{N} items of various shapes. For RU-DU resource assignment, the aim is to learn a generalised policy~\eqr{\pi(\cdot|s)} over various RU requirements. Consider the number of possible placement for items, searching for the sparse feasible solutions in large action space brings challenges to policy learning. In addition, the 2D BPP policy requires precise lookahead, since items cannot be withdrawn once placed.

Following the recent success of AlphaGo Zero~\cite{silver2017masteringAlphazero} and Expert Iteration~\cite{anthony2017thinkingfastandslow}, we apply the policy iteration and Monte-Carlo Tree Search (MCTS) on 2D BPP. A two-headed neural network \eqr{(\pi_\theta, v_\theta) = f_\theta (s)} takes the state representation as an input, with the output as a probability distribution over actions~\eqr{\pi_\theta(a|s)} and a state-value estimation~\eqr{v_\theta(s)}. For each state \eqr{s}, \eqr{M} MCTS simulations are performed, guided by \eqr{f_\theta}. The outputs of MCTS simulations are refined action probability distribution~\eqr{\hat{\pi}(a|s)} and state-value estimations~\eqr{\hat{v}(s)}. Taking the outputs of MCTS simulations as target policy and target state-value, \eqr{f_\theta} is trained in a supervised manner, minimising the following loss function:
\begin{equation}
\label{eq:loss func}
    l = (v_\theta (s) - \hat{v}(s))^2 - \hat{\pi}(a|s)\log \pi_\theta (a|s)
\end{equation}

We apply Ranked Reward (R2) as a reward-reshaping method to improve the MCTS search by self-play, presenting an adversary to the current agent from its own past performance~\cite{laterre2018ranked}. R2 maintains a fixed-length reward buffer ~\eqr{\mathcal{B}} with recent MDP final rewards. A threshold \eqr{r_\alpha} is used for reward-reshaping, which is the \eqr{\alpha^{\text{th}}} percentile of the reward buffer. As presented in~\cite{laterre2018ranked}, the MDP reward~\eqr{r} is reshaped to ranked reward~\eqr{z} as follows:
\begin{equation}
\label{eq:r2}
    z = 
    \begin{cases}
      1 & \text{if} \;\; r > r_{\alpha} \;\; \text{or} \;\; r = r_{\max} \\
      -1 & \text{if} \;\; r < r_{\alpha} \\
      \textit{random}(1, -1) & \text{if} \;\; r = r_{\alpha}\;\; \text{and} \;\; r < r_{\max}
    \end{cases}
\end{equation}
Here \eqr{\textit{random}(a, b)} performs a random selection between \eqr{a} and \eqr{b} with equal probabilities. \eqr{r_{\max}} is the upper bound of MDP reward, i.e., \eqr{1.0} in the 2D BPP. Algorithm~\ref{al:bpp-r2} shows the 2D BPP algorithm with ranked reward inspired self-play. Note that in this problem, rewards in~\eqr{\mathcal{B}} are from different instances of 2D BPP, which prevents the policy from overfitting to some of the instances. After each training iteration, the sample buffer~\eqr{\mathcal{D}} is cleared to stabilize the training process, because new rewards have been added to \eqr{\mathcal{B}}, resulting in the change of~\eqr{\alpha^{\text{th}}} percentile.

\begin{algorithm}[htp]
\SetAlgoLined
    \textbf{Input} Bin packing problem parameters; a percentile \eqr{\alpha}  \\
    \textbf{Output} Trained neural network \eqr{f_\theta}  \\
    Initialize neural network parameters \eqr{\theta}\\ 
    Initialize sample buffer \eqr{\mathcal{D}} and reward buffer \eqr{\mathcal{B}} and \eqr{\mathcal{B}^\prime}. Set \eqr{\mathcal{B}^\prime=\mathcal{B}} \\

\For{training iteration = \eqr{1,...,K}}{
\For{\eqr{episode = 1,...,J}}{
Generate a new instance of 2D BPP, with items \eqr{\mathcal{I}}

Initialize the bin as a zero occupancy plane

\For{step = \eqr{1,..., N}}{
Perform a Monte-Carlo Tree Search guided by \eqr{f_\theta}

Sample \eqr{a_t} from MCTS-improved policy \eqr{\pi(\cdot|s_t)}

Get \eqr{y_t} using Algorithm~\ref{algo:find y_t given a_t}

Perform action <\eqr{a_t, y_t}>, get the next state \eqr{s_{t+1}} and reward \eqr{r_t}}

Save \eqr{r_N} in \eqr{\mathcal{B}^\prime}

Get ranked reward \eqr{z} according to Eq.~\ref{eq:r2} using~\eqr{\mathcal{B}}

Store \eqr{(s_t, \pi(\cdot|s_t), z)} for all steps in \eqr{\mathcal{D}}
}
\For{step = \eqr{0,..., \tau}}{
Sample mini-batch \eqr{d} from \eqr{\mathcal{D}}

Update \eqr{\theta} by minimising Eq.~\ref{eq:loss func} on \eqr{d}
}

Clear \eqr{\mathcal{D}}

\eqr{\mathcal{B} = \mathcal{B}^\prime}
}
\caption{2D BPP with ranked reward inspired self-play}
\label{al:bpp-r2}
\end{algorithm}

\begin{algorithm}[htb]
\SetAlgoLined
    \textbf{Input} \eqr{a_t = <\mathcal{I}_t, x_t>}; \eqr{I_t = (w, h)}; Bin occupancy state \eqr{S^{\text{bin}}}.\\
    \textbf{Output} \eqr{y_t} \\
    Initialize \eqr{y_t = \emptyset}; \eqr{i = 0}; \eqr{y^\prime = 0}
    
    \While{i < h}{
    
        \For{y=\eqr{y^\prime+1,y^\prime+2,...}}{
              \eIf{\eqr{\sum S^{\text{bin}}[x_t:x_t+w-1, y] = 0} }{
                
                \eqr{y_t = y_t \cup [y]}
                
                \eqr{i = i+1}
                
                \eqr{y^\prime = y}
                
                \textbf{break}
               }{\textbf{continue}}
              }
    }     

\caption{Physical resource allocation for action \eqr{a_t}}
\label{algo:find y_t given a_t}
\end{algorithm}

\section{Experiments and Results}
\label{sec: experiments}
In this section, we first introduce the neural network architecture used in the 2D BPP. To validate the self-play learning approach, we create a representative 2D bin packing environment with 10 items, in which instances are generated by randomly slicing the bin. A policy~\eqr{f_\theta} is trained on the representative 2D bin packing environment. We then apply the trained policy to a dataset derived from a live network covering the central area of Bristol, UK, to optimise the RU-DU resource assignment.

\subsection{Neural Network Architecture}
The objective of the 2D BPP in this work is to minimise the height of the bin without setting an initial limitation on the bin size. To make the neural network model applicable, we set ``virtual'' bin width~\eqr{W^\prime} and height~\eqr{H^\prime}. Apparently, \eqr{W^\prime = W^*}, defined by the latency requirement. The neural network takes as input a \eqr{H^\prime \times W^\prime \times (N+1)} dimensional tensor, where~\eqr{N} is the number of items. We apply the neural network structure as presented in~\cite{espeholt2018impala}, with 15 convolutional layers. For the policy head, a softmax fully connected layer is applied to produce a probabilities distribution over all possible actions, while the value head applies another fully connect layer to estimate the continuous state-value.

\subsection{Model Training}
\begin{figure*}[htp]
    \centering
    \begin{subfigure}[c]{0.4\textwidth}
        \includegraphics[width = 2.2 in]{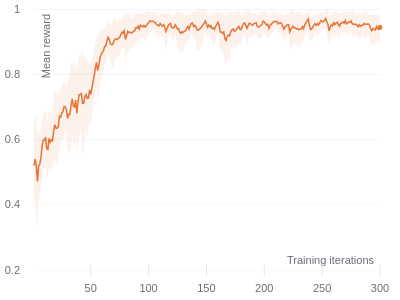}
        \caption{Mean reward.}
        \label{fig:mean reward}
    \end{subfigure}
    \hspace{0.6 cm}
    \begin{subfigure}[c]{0.4\textwidth}
        \includegraphics[width = 2.2 in]{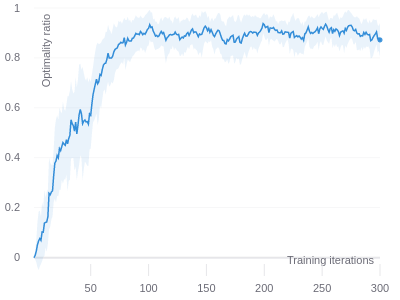}
        \caption{Optimality ratio.}
        \label{fig:optim percent}
    \end{subfigure}
\caption{Mean rewards and optimality ratio of 2D BPP during 300 training iterations.}
\label{fig: training reward and optim percent}
\end{figure*}

In each iteration, 20 different instances are generated from the representative 2D bin packing environment, by slicing a bin with size~\eqr{W^*, H^*} into 10 items. Here we set~\eqr{W^\prime, H^\prime = 15}, \eqr{W^* = 15}, and~\eqr{H^*} can be any integer between \textit{2} to \textit{15}, to train a policy applicable for different RU requirements. Note that we only train one policy \eqr{f_\theta} for varying values of \eqr{H^*}. Self-play results for all instances are stored in~\eqr{\mathcal{B}} with length 100. At every self-play step, 200 MCTS simulations are conducted to generate improved policy and state-value estimation. For the ranked reward, we set~\eqr{\alpha = 75} as suggested in~\cite{laterre2018ranked}. Fig.~\ref{fig: training reward and optim percent} shows the mean reward and the ratio of ``optimal'' packing during training, i.e., the ratio of bin packing results where \eqr{\tilde{H} = H^*}.

\begin{figure}[h]
    \centering
    \begin{subfigure}[c]{0.5\textwidth}
        \includegraphics[width = 4 in]{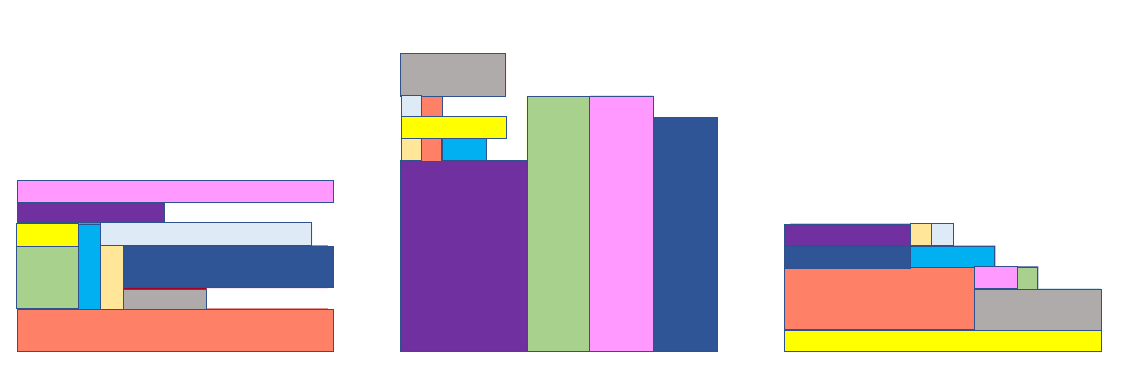}
        \caption{Results of the HVRAA method.}
        \label{fig:bpp_hvraa}
    \end{subfigure} \\
    \begin{subfigure}[c]{0.5\textwidth}
        \includegraphics[width = 4 in]{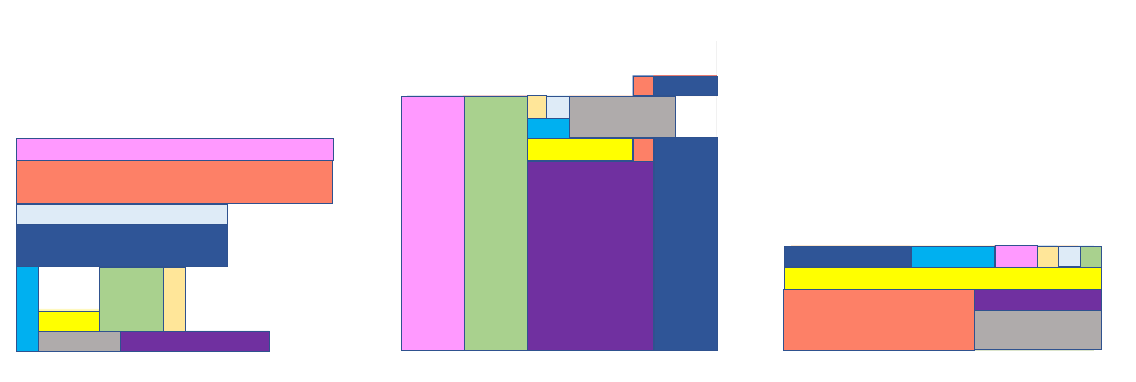}
        \caption{Results of the Lego heuristic method.}
        \label{fig:bpp_lego}
    \end{subfigure}\\
    \begin{subfigure}[c]{0.5\textwidth}
        \includegraphics[width = 4.0 in]{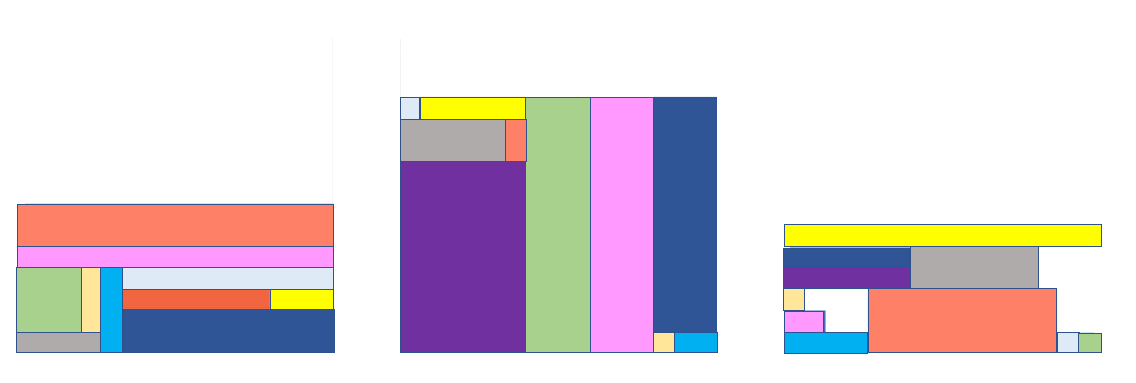}
        \caption{Results of the MCTS method.}
        \label{fig:bpp_mcts}
    \end{subfigure} \\
    \begin{subfigure}[c]{0.5\textwidth}
        \includegraphics[width = 4.0 in]{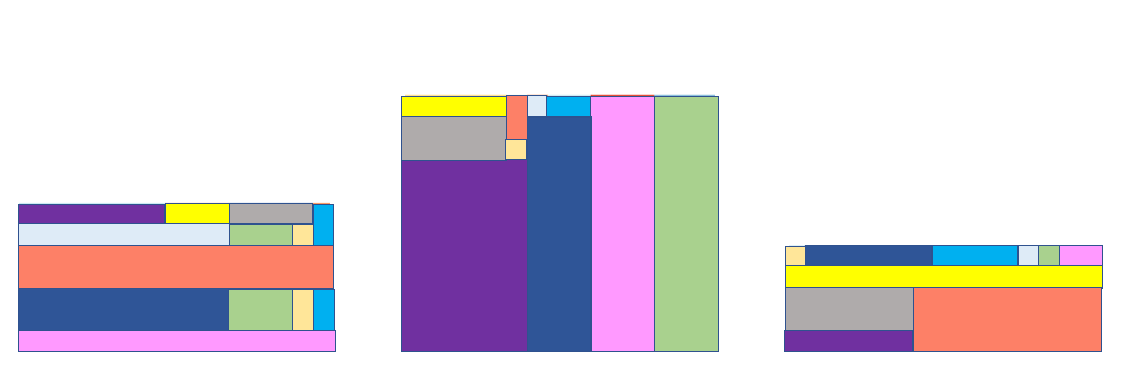}
        \caption{Results of the self-play learning method.}
        \label{fig:bpp_self_play}
    \end{subfigure}
\caption{Results of three 2D BPP instances using baseline methods and the proposed self-play learning strategy. For each instance, the same colour represents the same item.}
\label{fig: standard bpp test}
\end{figure}

We compare the self-play learning strategy with a set of baselines: a heuristic virtual resource allocation algorithm (HVRAA) proposed in~\cite{zhu2017threeHVRAA}; the Lego heuristics for bin packing~\cite{hu2018multiLego}; and the vanilla MCTS using Monte-Carlo rollout. The test is conducted on 100 instances of 10-item 2D BPP, with randomly sampled \eqr{H^*} from~\eqr{U(2,15)} and randomly generated item shapes. Table~\ref{tab:2D bpp test standard} shows the average reward~\eqr{\Bar{r}}, standard deviation of rewards~\eqr{\sigma_r} and the optimality ratio~\eqr{R_{\tilde{H}=H^*}} on the test set. The proposed self-play learning method outperforms the baseline methods on the average reward, with the second lowest standard deviation. The optimality ratio of the self-play learning approach reaches 94\%, which is 29\% higher than the second highest value. This proves the effectiveness and the robustness of the trained model across various bin packing problems, with items of different sizes. Fig.~\ref{fig: standard bpp test} visualizes three instances of 2D BPP solutions by baseline methods and the proposed self-play learning method, where items are initially generated by slicing bins with ~\eqr{H^* = [7, 12, 5]}, respectively.

\begin{table}[htb]
    \centering
    \begin{tabular}{c|c|c|c}
    \hline
    & \eqr{\Bar{r}} & \eqr{\sigma_r} & \eqr{R_{\tilde{H}=H^*}} \\
    \hline
        HVRAA~\cite{zhu2017threeHVRAA} &  0.896 & 0.239 & 0.65\\
        Lego heuristics~\cite{hu2018multiLego} & 0.737 & 0.349 & 0.44\\
        MCTS & 0.936 & \textbf{0.097} & 0.62\\ 
        Self-play learning method & \textbf{0.964} & 0.160 & \textbf{0.94} \\
    \hline
    \end{tabular}
    \caption{Performance on 2D BPP for the proposed self-play learning method and other baseline methods.}
    \label{tab:2D bpp test standard}
\end{table}

\subsection{Experiments on Real Sites Data}
\begin{figure}[ht]
\begin{minipage}[b][][b]{0.45\textwidth}
\centering
 \includegraphics[width=2.0 in]{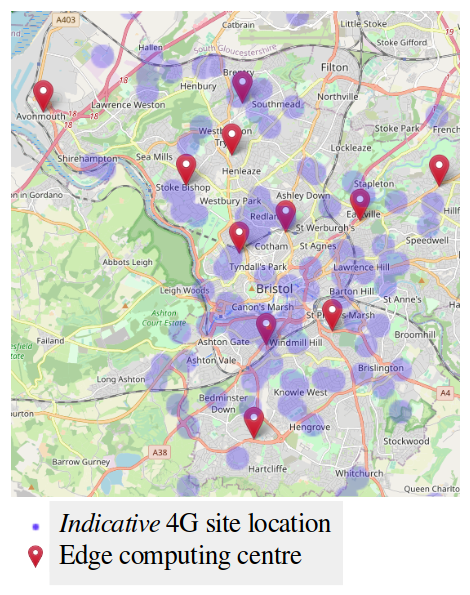}
 \caption{Potential deployment area for ORAN in Bristol, UK.}
 \label{fig:bristol_map}
\end{minipage}
\hspace{0.5 cm}
\begin{minipage}[b][][b]{0.45\textwidth}
 \centering
 \includegraphics[width=2.4 in]{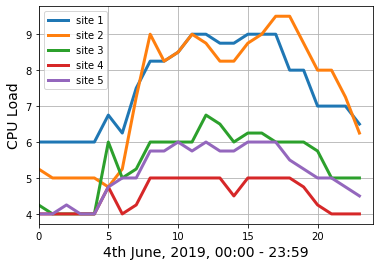}
 \caption{The 24-hour mean CPU load of five 4G sites in Bristol, UK.}\label{fig:sites cpu load}
\end{minipage}
\end{figure}

We select a~\eqr{14 \times 13} \SI{}{\km \squared} area in Bristol, UK as the potential ORAN deployment area which covers the city centre. It contains more than 100 ``4G'' sites, currently hosting a mix of 4G and 5G base stations, and more than 10 viable locations for edge computing centres. Fig.~\ref{fig:bristol_map} shows the map of this area. Sites and potential edge computing centre data is provided by BT\footnote{https://www.bt.com/}. In the ORAN structure, current sites can be seen as RUs, while DUs are allocated in edge computing centres\footnote{This work is looking at future RAN structure. We use viable locations for edge computing centres as potential DUs in this work. The feasibility of hosting DUs in these locations has not yet been validated by the mobile network operator.}. In this work, we perform RU-DU resource assignment on 2 DUs, located at two different postcode areas. For the physical resource, we focus on the CPU.


The latency between RU and DU can be calculated as
\begin{equation}
    t = F \times \frac{d}{c}
\end{equation}
where \eqr{d} is the straight line distance between the DU and RU, \eqr{c} is the speed of light, and \eqr{F} is a reduction factor, due to the speed reduction in the fibre and the fact that fibre cables are not straight. Empirically, we set~\eqr{F = 2.5}. The latency requirement for Ethernet mobile fronthaul is \SI{100}{\micro \s}~\cite{ericsson2015common}. For each DU, it is assumed to be connected with 10 nearest RUs through fronthaul network, which guarantees low latency for all RU-DU connections.

Fig.~\ref{fig:sites cpu load} shows the hourly average CPU requirements for 5 RUs on 4th June 2019, randomly selected from the Bristol map. Peak CPU requirements always occur at around 17:00 and the off-peak hours are from 21:00 to 4:00. We assume the CPU resource requirements are integers sampled from a uniform distribution~\eqr{\mathcal{U}(\mu-\delta, \mu+\delta)}, where~\eqr{\mu} and ~\eqr{\mu+\delta} are the mean and max CPU requirements for one site. Required processing time units are sampled from~\eqr{\mathcal{U}(1, T^\prime)}, where~\eqr{T^\prime} is the maximal allowed processing time.

In this work, we focus on peak hours. Different from the representative 2D BPP, there is no guarantee on the existence of a bin~\eqr{(W^*, H^*)} whose size equals to the total area of items, i.e., for the reward, ~\eqr{r_{\max} = 1} does not always hold for real sites data. We introduce a new metric, the resource utilisation, for evaluating the resource assignment performance. The resource utilisation is defined as
\begin{equation}
    U = \frac{\sum_{i=1}^N w_i * h_i}{\tilde{W} * \tilde{H}}
\end{equation}

\begin{figure}[h]
    \centering
    \begin{subfigure}[c]{0.5\textwidth}
        \includegraphics[width = 3.5 in]{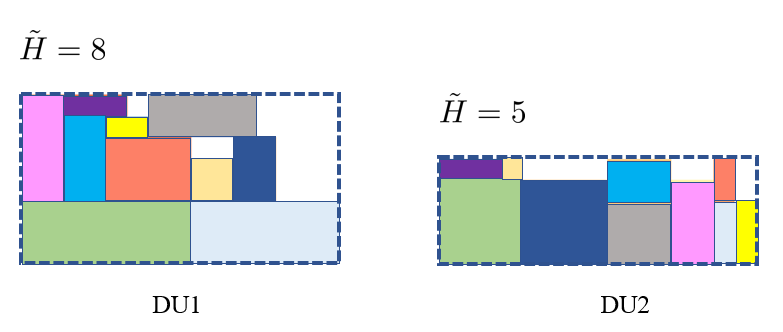}
        \caption{Results of the HVRAA method.}
        \label{fig:sites_hvraa}
    \end{subfigure}\\
    \begin{subfigure}[c]{0.5\textwidth}
        \includegraphics[width = 3.5 in]{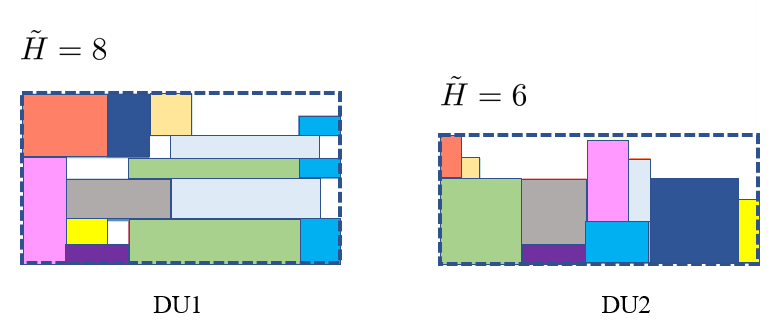}
        \caption{Results of the Lego heuristics method.}
        \label{fig:sites_lego}
    \end{subfigure}\\
    \begin{subfigure}[c]{0.5\textwidth}
        \includegraphics[width = 3.5 in]{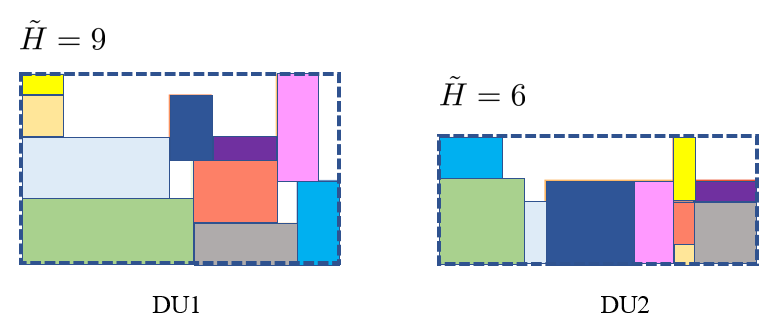}
        \caption{Results of the MCTS method.}
        \label{fig:sites_mcts}
    \end{subfigure} \\
    \begin{subfigure}[c]{0.5\textwidth}
        \includegraphics[width = 3.5 in]{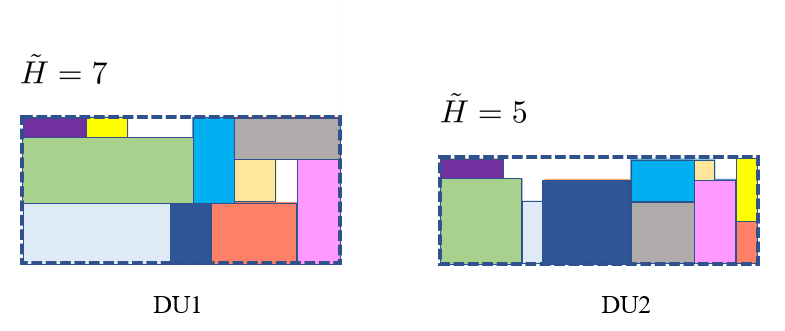}
        \caption{Results of the self-play learning method.}
        \label{fig:sites_self_play}
    \end{subfigure}
\caption{RU-DU resource assignment results of two instances for DU1 and DU2. Dashed lines show optimized bins \eqr{(\tilde{W}, \tilde{H})} of each problem. Same with Fig.~\ref{fig: standard bpp test}, for each instance, the same colour represents the same item.}
\label{fig: RU-DU bpp results}
\end{figure}

For each DU, we test on 10 instances of peak hour RU requirements. Table~\ref{tab:real sites bpp} shows the average reward~\eqr{\Bar{r}} and the average utilisation \eqr{\Bar{U}} of DU1 and DU2, and Fig.~\ref{fig: RU-DU bpp results} visualises two instances. Note that the optimized bin sizes are marked by dashed lines in Fig.~\ref{fig: RU-DU bpp results}. From Table \ref{tab:real sites bpp} we see that the proposed self-play learning method has the highest average reward for both DUs, with the average utilisation around \eqr{86\%}. Although the neural network model is trained on 2D BPPs with the guaranteed existence of \eqr{r_{\max} = 1}, it can be applied on instances that are different from the representative 2D BPPs, achieving the best possible packing solution with minimal bin height. This experiment shows that the model trained in an offline manner can be utilised for dynamic resource assignment, without the need for online interactions or demonstrator data. Using offline training to replace online interaction can avoid querying the live network, reducing the time and query signal cost in network operation. Model re-use can greatly lower the training time and resource cost for network management, as well as reducing the potential operational latency.

\begin{table}[htb]
    \centering
    \begin{tabular}{c|c|c|c|c}
    \hline
    & \multicolumn{2}{c|}{DU1} & \multicolumn{2}{c}{DU2}\\
    \hline
    &\eqr{\Bar{r}} & \eqr{\Bar{U}} & \eqr{\Bar{r}} & \eqr{\Bar{U}} \\
    \hline
    HVRAA &  0.879 & 0.809 & 0.938 & 0.810\\
    Lego heuristics  & 0.841 & 0.774 & 0.811 & 0.700 \\
    MCTS & 0.847 & 0.781 & 0.847 & 0.732\\
    Self-play learning method & \textbf{0.943} & \textbf{0.869} & \textbf{1.000} & \textbf{0.864} \\
    \hline
    \end{tabular}
    \caption{Resource assignment performance on real sites.}
    \label{tab:real sites bpp}
\end{table}

\section{Conclusion and Future Work}
\label{sec: discussion}
This work studied the RU-DU resource assignment problem under the ORAN dis-aggregated structure. RU requirements are modelled as 2D items, with the objective of finding a bin with minimal height to pack all items, given a restriction on the bin width. A self-play reinforcement learning strategy is applied to the 2D BPP. The use of ranked reward enables the RL agent to compete against itself, forming a self-play learning structure. Trained on representative BPP environments where the existence of optimal solutions is guaranteed, the model can be applied to bin packing problems outside the original assumption. This would help maintain compact utilisation of DU resources, achieving cost-effective RAN operations. The offline training and well-generalize model reduce both training resource cost and latency for network operations. 


The RU-DU resource assignment with multiple resources remains an open problem. DUs have different capacities for various resources types. One potential way to tackle this problem is to model the multi-resource assignment as a multi-agent RL problem, with each agent dedicated to one resource type. In multi-agent RL, multiple agents can be trained to work in a cooperative way to assign DU resources for RU requirements. Not only is this challenging to RL but also one of the key issues to deploy such algorithms in ORAN. 

Another challenge is brought by the scale of RAN. The densification of network leads to large-scale RU-DU distribution. Consider the 2D BPP model presented in this paper, the increasing size of action space will become a bottleneck with a large scale RAN. Besides, as the model complicates or the model scale enlarges, the cost and latency brought by running trained models on the network could be non-negligible.

\section*{Acknowledgement}
This work is funded by the Next-Generation Converged Digital Infrastructure (NG-CDI) Project, supported by BT and Engineering and Physical Sciences Research Council (EPSRC), Grant ref. EP/R004935/1.


\medskip

\small

\bibliographystyle{unsrt}
\bibliography{ref}

\begin{thebibliography}{10}

\bibitem{Nokiawhitepaper}
Deploying 5{G} networks, 2020, {N}okia corporation. {A}vailable at:
  https://www.nokia.com/networks/5g/mo\\bile/5g-resources/.

\bibitem{singh2020evolution}
Sameer~Kumar Singh, Rohit Singh, and Brijesh Kumbhani.
\newblock The evolution of radio access network towards open-ran: Challenges
  and opportunities.
\newblock In {\em 2020 IEEE Wireless Communications and Networking Conference
  Workshops (WCNCW)}, pages 1--6. IEEE, 2020.

\bibitem{ren2020resource}
Hong Ren, Cunhua Pan, Yansha Deng, Maged Elkashlan, and Arumugam Nallanathan.
\newblock Resource allocation for secure urllc in mission-critical iot
  scenarios.
\newblock {\em IEEE Transactions on Communications}, 68(9):5793--5807, 2020.

\bibitem{hussain2020machine}
Fatima Hussain, Syed~Ali Hassan, Rasheed Hussain, and Ekram Hossain.
\newblock Machine learning for resource management in cellular and iot
  networks: Potentials, current solutions, and open challenges.
\newblock {\em IEEE Communications Surveys \& Tutorials}, 22(2):1251--1275,
  2020.

\bibitem{samsungwhitepaper}
{O}pen {RAN} - {T}he open road to 5{G}, 2017, {SAMSUNG}. {A}vailable at:
  https://image-us.samsung.com/{S}ams\\ungus/samsungbusiness/pdfs/{O}pen-{RAN}-{T}he-{O}pen-{R}oad-to-5{G}.pdf.

\bibitem{oranwhitepaper}
\uppercase{O-RAN}: {T}owards an {O}pen and {S}mart {RAN}, {O-RAN} {A}lliance
  {W}hite {P}aper, {O}ctober 2018, {O-RAN} {A}lliance. {A}vaiable at:
  https://static1.squarespace.com/static/5ad774cce74940d7115044b0/t/5bc79b371905f4197055e\\8c6/1539808057078/{O-RAN+WP+FI}nal+181017.pdf.

\bibitem{3gppr14}
Technical specification group radio access network: Study on new radio access
  technology: Radio access architecture and interfaces (release 14), 3{GPP}
  {TR} 38.801 v14.0.0, 2017, March 2017.

\bibitem{yi2018comprehensiveNFV}
Bo~Yi, Xingwei Wang, Keqin Li, Min Huang, et~al.
\newblock A comprehensive survey of network function virtualization.
\newblock {\em Computer Networks}, 133:212--262, 2018.

\bibitem{wu2015cloud}
Jun Wu, Zhifeng Zhang, Yu~Hong, and Yonggang Wen.
\newblock Cloud radio access network ({C-RAN}): a primer.
\newblock {\em IEEE Network}, 29(1):35--41, 2015.

\bibitem{parallelwirelesswhitepaper}
{P}arallel wireless creates {OpenRAN} “{A}ll {G}” radio access network
  architecture, {J}une 2020, \uppercase{P}arallel wireless white paper.
  {A}vailabel at:
  https://www.parallelwireless.com/wp-content/uploads/parallel-wireless-creates-openran-all-g-radio-access-network.pdf.

\bibitem{tang2019systematic}
Jianhua Tang, Tony~QS Quek, Tsung-Hui Chang, and Byonghyo Shim.
\newblock Systematic resource allocation in cloud {RAN} with caching as a
  service under two timescales.
\newblock {\em IEEE Transactions on Communications}, 67(11):7755--7770, 2019.

\bibitem{yang2020survivableheu}
Song Yang, Nan He, Fan Li, Stojan Trajanovski, Xu~Chen, Yu~Wang, and Xiaoming
  Fu.
\newblock Survivable task allocation in cloud radio access networks with mobile
  edge computing.
\newblock {\em IEEE Internet of Things Journal}, 2020.

\bibitem{xia2019programmable}
Wenchao Xia, Tony~QS Quek, Jun Zhang, Shi Jin, and Hongbo Zhu.
\newblock Programmable hierarchical {C-RAN}: From task scheduling to resource
  allocation.
\newblock {\em IEEE Transactions on Wireless Communications}, 18(3):2003--2016,
  2019.

\bibitem{mharsi2019cloud}
Niezi Mharsi.
\newblock {\em Cloud-Radio Access Networks: design, optimization and
  algorithms}.
\newblock PhD thesis, 2019.

\bibitem{silver2017masteringAlphazero}
David Silver, Julian Schrittwieser, Karen Simonyan, Ioannis Antonoglou, Aja
  Huang, Arthur Guez, Thomas Hubert, Lucas Baker, Matthew Lai, Adrian Bolton,
  et~al.
\newblock Mastering the game of {G}o without human knowledge.
\newblock {\em nature}, 550(7676):354--359, 2017.

\bibitem{silver2017masteringchess}
David Silver, Thomas Hubert, Julian Schrittwieser, Ioannis Antonoglou, Matthew
  Lai, Arthur Guez, Marc Lanctot, Laurent Sifre, Dharshan Kumaran, Thore
  Graepel, et~al.
\newblock Mastering chess and shogi by self-play with a general reinforcement
  learning algorithm.
\newblock {\em arXiv preprint arXiv:1712.01815}, 2017.

\bibitem{mnih2015humanlevelcontrol}
Volodymyr Mnih, Koray Kavukcuoglu, David Silver, Andrei~A Rusu, Joel Veness,
  Marc~G Bellemare, Alex Graves, Martin Riedmiller, Andreas~K Fidjeland, Georg
  Ostrovski, et~al.
\newblock Human-level control through deep reinforcement learning.
\newblock {\em nature}, 518(7540):529--533, 2015.

\bibitem{vinyals2019grandmasterSCII}
Oriol Vinyals, Igor Babuschkin, Wojciech~M Czarnecki, Micha{\"e}l Mathieu,
  Andrew Dudzik, Junyoung Chung, David~H Choi, Richard Powell, Timo Ewalds,
  Petko Georgiev, et~al.
\newblock Grandmaster level in {S}tar{C}raft {II} using multi-agent
  reinforcement learning.
\newblock {\em Nature}, 575(7782):350--354, 2019.

\bibitem{luong2019applications}
Nguyen~Cong Luong, Dinh~Thai Hoang, Shimin Gong, Dusit Niyato, Ping Wang,
  Ying-Chang Liang, and Dong~In Kim.
\newblock Applications of deep reinforcement learning in communications and
  networking: A survey.
\newblock {\em IEEE Communications Surveys \& Tutorials}, 21(4):3133--3174,
  2019.

\bibitem{zhao2018deepD3QN}
Nan Zhao, Ying-Chang Liang, Dusit Niyato, Yiyang Pei, and Yunhao Jiang.
\newblock Deep reinforcement learning for user association and resource
  allocation in heterogeneous networks.
\newblock In {\em 2018 IEEE Global Communications Conference (GLOBECOM)}, pages
  1--6. IEEE, 2018.

\bibitem{liu2019deepMEC}
Yi~Liu, Huimin Yu, Shengli Xie, and Yan Zhang.
\newblock Deep reinforcement learning for offloading and resource allocation in
  vehicle edge computing and networks.
\newblock {\em IEEE Transactions on Vehicular Technology}, 68(11):11158--11168,
  2019.

\bibitem{wang2019smartMEC}
Jiadai Wang, Lei Zhao, Jiajia Liu, and Nei Kato.
\newblock Smart resource allocation for mobile edge computing: A deep
  reinforcement learning approach.
\newblock {\em IEEE Transactions on emerging topics in computing}, 2019.

\bibitem{qi2019deepslicing}
Chen Qi, Yuxiu Hua, Rongpeng Li, Zhifeng Zhao, and Honggang Zhang.
\newblock Deep reinforcement learning with discrete normalized advantage
  functions for resource management in network slicing.
\newblock {\em IEEE Communications Letters}, 23(8):1337--1341, 2019.

\bibitem{wang2019dataslicing}
Haozhe Wang, Yulei Wu, Geyong Min, Jie Xu, and Pengcheng Tang.
\newblock Data-driven dynamic resource scheduling for network slicing: A deep
  reinforcement learning approach.
\newblock {\em Information Sciences}, 498:106--116, 2019.

\bibitem{liang2019spectrum}
Le~Liang, Hao Ye, and Geoffrey~Ye Li.
\newblock Spectrum sharing in vehicular networks based on multi-agent
  reinforcement learning.
\newblock {\em IEEE Journal on Selected Areas in Communications},
  37(10):2282--2292, 2019.

\bibitem{lin2020dynamicspectrum}
Yun Lin, Meiyu Wang, Xianglong Zhou, Guoru Ding, and Shiwen Mao.
\newblock Dynamic spectrum interaction of {UAV} flight formation communication
  with priority: A deep reinforcement learning approach.
\newblock {\em IEEE Transactions on Cognitive Communications and Networking},
  2020.

\bibitem{bengio2020machinelearning}
Yoshua Bengio, Andrea Lodi, and Antoine Prouvost.
\newblock Machine learning for combinatorial optimization: a methodological
  tour d’horizon.
\newblock {\em European Journal of Operational Research}, 2020.

\bibitem{mazyavkina2020reinforcementsurvey}
Nina Mazyavkina, Sergey Sviridov, Sergei Ivanov, and Evgeny Burnaev.
\newblock Reinforcement learning for combinatorial optimization: A survey.
\newblock {\em arXiv preprint arXiv:2003.03600}, 2020.

\bibitem{laterre2018ranked}
Alexandre Laterre, Yunguan Fu, Mohamed~Khalil Jabri, Alain-Sam Cohen, David
  Kas, Karl Hajjar, Torbjorn~S Dahl, Amine Kerkeni, and Karim Beguir.
\newblock Ranked reward: Enabling self-play reinforcement learning for
  combinatorial optimization.
\newblock {\em arXiv preprint arXiv:1807.01672}, 2018.

\bibitem{wang2020tacklingMS}
Hui Wang, Mike Preuss, Michael Emmerich, and Aske Plaat.
\newblock Tackling {M}orpion {S}olitaire with {A}lpha{Z}ero-like ranked reward
  reinforcement learning.
\newblock {\em arXiv preprint arXiv:2006.07970}, 2020.

\bibitem{xu2019learning}
Ruiyang Xu and Karl Lieberherr.
\newblock Learning self-game-play agents for combinatorial optimization
  problems.
\newblock {\em arXiv preprint arXiv:1903.03674}, 2019.

\bibitem{abe2019solvingNPHard}
Kenshin Abe, Issei Sato, and Masashi Sugiyama.
\newblock Solving {NP}-hard problems on graphs by reinforcement learning
  without domain knowledge.
\newblock {\em Simulation}, 1:1--1, 2019.

\bibitem{larsen2018surveysplit}
Line~MP Larsen, Aleksandra Checko, and Henrik~L Christiansen.
\newblock A survey of the functional splits proposed for 5{G} mobile crosshaul
  networks.
\newblock {\em IEEE Communications Surveys \& Tutorials}, 21(1):146--172, 2018.

\bibitem{anthony2017thinkingfastandslow}
Thomas Anthony, Zheng Tian, and David Barber.
\newblock Thinking fast and slow with deep learning and tree search.
\newblock In {\em Advances in Neural Information Processing Systems}, pages
  5360--5370, 2017.

\bibitem{espeholt2018impala}
Lasse Espeholt, Hubert Soyer, Remi Munos, Karen Simonyan, Volodymir Mnih, Tom
  Ward, Yotam Doron, Vlad Firoiu, Tim Harley, Iain Dunning, et~al.
\newblock Impala: Scalable distributed deep-{RL} with importance weighted
  actor-learner architectures.
\newblock {\em arXiv preprint arXiv:1802.01561}, 2018.

\bibitem{zhu2017threeHVRAA}
Wei Zhu, Yi~Zhuang, and Long Zhang.
\newblock A three-dimensional virtual resource scheduling method for energy
  saving in cloud computing.
\newblock {\em Future Generation Computer Systems}, 69:66--74, 2017.

\bibitem{hu2018multiLego}
Haoyuan Hu, Lu~Duan, Xiaodong Zhang, Yinghui Xu, and Jiangwen Wei.
\newblock A multi-task selected learning approach for solving new type 3d bin
  packing problem.
\newblock {\em arXiv preprint arXiv:1804.06896}, 2018.

\bibitem{ericsson2015common}
AB~Ericsson et~al.
\newblock Common public radio interface ({CPRI}) specification v7.0, 2015.

\end{thebibliography}

\end{document}